\documentclass[aps,prc,superscriptaddress,prc,twocolumn,showpacs,amsmath,amssymb]{revtex4-1}
\usepackage{amsmath}
\usepackage{amssymb}
\usepackage{dcolumn}
\usepackage{bm}
\usepackage{graphicx}
\usepackage{color}   
\usepackage[varg]{txfonts} 
\usepackage[latin1]{inputenc}
\graphicspath{{Figures/}}
\usepackage[hidelinks]{hyperref}

\begin{document}

\title{Evaporation Channel as a Tool to Study Fission Dynamics}

\author{A.~Di~Nitto}\email[corresponding author e-mail: ]{a.dinitto@gsi.de}\affiliation{Istituto Nazionale di Fisica Nucleare, Sezione di Napoli, 80126 Napoli, Italy}\affiliation{Johannes Gutenberg-Universit{\"a}t Mainz, 55099 Mainz, Germany}
\author{E.~Vardaci}\affiliation{Istituto Nazionale di Fisica Nucleare, Sezione di Napoli, 80126 Napoli, Italy} \affiliation{Dipartimento di Fisica, Universit\'{a} degli Studi di Napoli ``Federico II'', Italy}
\author{G.~La Rana}\affiliation{Istituto Nazionale di Fisica Nucleare, Sezione di Napoli, 80126 Napoli, Italy} \affiliation{Dipartimento di Fisica, Universit\'{a} degli Studi di Napoli ``Federico II'', Italy}
\author{P.~N.~Nadtochy}\affiliation{Istituto Nazionale di Fisica Nucleare, Sezione di Napoli, 80126 Napoli, Italy} \affiliation{ Omsk State Technical University, Mira prospekt 11, 644050 Omsk, Russia}
\author{G.~Prete}\affiliation{Laboratori Nazionali di Legnaro dell'Istituto Nazionale di Fisica Nucleare, 35020 Legnaro (Padova), Italy}
\date{\today}

\begin{abstract}
The dynamics of the fission process is expected to affect the evaporation residue 
cross section because of the fission hindrance due to the nuclear viscosity. 
Systems of intermediate fissility constitute a suitable environment for testing 
such hypothesis, since they are characterized by evaporation residue cross sections 
comparable or larger than the fission ones.
Observables related to emitted charged particle, due to their relatively high emission 
probability, can be used to put stringent constraints on models describing the 
excited nucleus decay and to recognize the effects of fission dynamics.
In this work model simulations are compared with the experimental data collected via
the $\rm ^{32}S + ^{100}Mo$ reaction at $\rm E_{lab}$= 200\,MeV.
By comparing an extended set of evaporation channel observables the limits of the 
statistical model and the large improvement coming by using a dynamical model are evidenced.
The importance of using a large angular covering apparatus to extract 
the observable is stressed.
The opportunity to measure more sensitive observables by a new detection device 
in operation at LNL are also discussed.
\end{abstract}


\maketitle
\section{Introduction}
Since the discovery of nuclear fission in 1939 \cite{Meit39,Hahn39} a large effort was 
devoted to provide a realistic description of this complex phenomenon originated by 
the interplay of macroscopic and microscopic degrees of freedom in a nucleus.
With the advent of heavy-ion accelerators the study of fission was extended to
a new variety of nuclei produced in few-nucleon direct transfer reactions \cite{Saxe02,Legu16} 
or in complete fusion reactions (as in recent works \cite{Khuy15,Kozu16}).
More recently by using reactions with radioactive beams it was demonstrated that 
fission studies can be performed also on very neutron-deficient mercury-to-thorium 
nuclei \cite{Andr13}.

It is well established that fission is a slow process dominated by nuclear viscosity. 
A very striking experimental evidence of this behaviour is the excess of pre-scission 
light particles, with respect to the predictions of the statistical model (SM), and 
its dependence on the excitation energy \cite{Hind92}.
Phenomenological studies based on the SM predictions were carried out with the aim to 
estimate the fission delay time, and, in some cases, extract the strength of nuclear 
viscosity. 
The estimates given by different authors predict a quite wide range of dissipation strengths 
and different dependencies on temperature and deformation (see reviews 
~\cite{Hils92, Frob98, VardPramana} and references therein).
However, this kind of approach is founded on the reliability of the SM to reproduce the 
observables in the evaporation residue (ER) channel, and this has not yet been fully explored.

The lack of experimental constraints to the model appears to be, in several cases, one of 
the source of controversial results.
By considering large set of observables the limits of the SM have been evidenced \cite{Dini11,Vard10}.
Dynamical models based on a stochastic approach combined with an evaporative model, for light 
particles and gamma quanta, seem to be a more suitable tool for the description of the 
collective evolution of nuclei \cite{Nadt10,Nadt16}.
Although much work has been devoted to fission dynamics, there are still many open questions:
the time-scale, the strength and nature of dissipation, as well as the dependence on the 
temperature and shape of the fissioning system are key items still to be disentangle.

The dynamics of the fission process is expected to affect the evaporation residue channel 
because of the fission hindrance due to nuclear viscosity.
Systems of intermediate fissility constitute a suitable environment for measuring potentially 
informative observables, being characterized by higher probability for 
charged particle emissions and integral ER cross section comparable with the fission one.
In order to address fission dynamics such advantages were largely exploited by using as probes 
the light particles \cite{Lara03} and, only recently, by using the fission-fragment charge 
distribution \cite{Mazu16}.
However in order to fruitfully benchmark the existing models the experimental uncertainties 
have to be minimized, thereby larger angular coverage apparatuses are essential to step forward.

Here we report on the measurement and analysis of the evaporation and fission decay of 
the compound nucleus $\rm ^{132}Ce$ at $\rm E_{x}$=122\,MeV, produced by the 200\,MeV $
\rm ^{32}S + ^{100}Mo$ reaction.
For this system ER and fusion fission (FF) angular distributions and cross sections, 
light charged particle (LCP) multiplicities and spectra as well as ER-LCP angular correlations 
were measured \cite{Dini11,Vard15}.
The measured quantities were compared with the SM calculations carried out by 
changing many physical ingredients of the model as well as with a dynamical model
based on the 3-D Langevin equations.
It was found that the ER observables, especially the LCP multiplicities and ER-LCP angular 
correlations, can be used not only to fix SM parameters, but also to provide constraints 
on the ingredients describing the fission mechanism. 
The analysis based on the dynamical model, i.e. considering a more realistic approach \cite{Vard15}, 
shows better data reproduction.
However there is still substantial room for improving the reliability of such conclusions. 
One mean is to investigate what observables can be still identified to be affected by fission 
dynamics.
Additional observables with such properties are for example the partial multiplicities and 
ER-LCP correlation angular distribution as function of evaporation residues angles that 
can be measured by coupling the new PPAC recently used at LNL ~\cite{Dini16} and the 
$\rm 8 \pi LP$ detection apparatus.

This article is organized as it follows. 
In Sec.~\ref{sec:Exp} we present in more detail the description of the 
experimental setup.
In Sec. ~\ref{sec:Mod} the theoretical models used for simulations are briefly introduced. 
In Sec.~\ref{sec:Res} we show the observables that provide benchmarks for   
parameters and prescriptions describing the fission dynamics.
We first discuss the result of comparison between experimental data and simulations and 
afterwards the advantages offered by using a recently developed detection setup.
In Sec.~\ref{sec:Con} we draw our conclusions.


\section{Experimental setup}
\label{sec:Exp}
The experiment was performed at the Tandem accelerator of Laboratori Nazionali 
di Legnaro. 
A pulsed beam of $\rm ^{32}$S of intensity of about 1-3\,enA was used to bombard 
a self-supporting $\rm ^{100}$Mo target 400 $\rm \mu g/cm^{2}$  thick. 
A beam burst with period of 800\,ns and duration of about 3\,ns was used.

\begin{figure}[h]
\centering
\includegraphics[width=9cm]{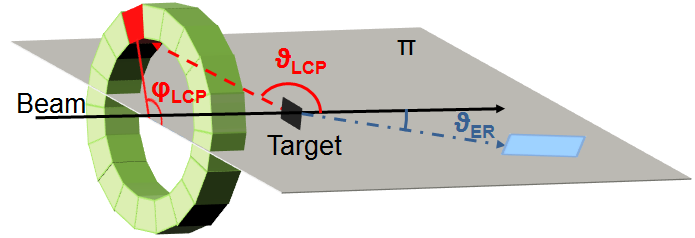}
\caption{(Color online) Schematic presentation of a coincident event where the LCP is detected 
by a Ball telescope and ER by a PPAC.
The emission angle of evaporation residue ($\rm \theta_{ER}$) is included between the beam 
direction and the line connecting the center of PPAC and the target. 
These two lines define 
the reaction plane $\rm \pi$.
The polar and azimuthal angles ($\rm \theta_{LCP}$) and ($\rm \phi_{LCP}$), respectively, 
define the direction of LCP impinging on a Ball telescope.
A Ring at ($\rm \theta_{LCP}$) is shown. 
It is made of 18 identical telescopes mounted at 
same polar angle and spanning all the possible azimuthal directions.}
\label{8PLP}
\end{figure}
We used the Ball sector of $\rm 8 \pi$LP apparatus \cite{Fior97} to detect light
charged particles and fission fragments, while the heavy residues were detected 
in a system of four parallel plate avalanche counters (PPAC) placed at forward 
angles.
The Ball, covering the polar angle from 34 to 165\,degrees, has a diameter of 
30\,cm and consists of 7 rings of $\rm \Delta E$-E telescopes placed co-axially around 
the beam direction. The schematic of one backward ring is shown in Fig.~\ref{8PLP}. 
Each ring contains 18 telescopes. 
Each telescope points towards the center of the target, and covers a polar angular opening 
of 17\,degrees.
Considering this geometry, the detectors in a ring have the same average polar angle 
with respect to the beam direction axis, and all together they cover the azimuthal angle 
from 0 to 360\,degrees. 
As a whole the Ball covers a solid angle of about $\rm 80 \%$ of $\rm 4 \pi$ by means of 
125 telescopes.
Each telescope is made by a first stage ($\rm \Delta E$) of 300\,$\rm \mu$m thick Si detectors 
followed by a second stage (E) of 5\,mm thick CsI(Tl) with photo-diode read out.
Particle identification was carried out by using the $\rm \Delta E-E$ technique for the 
particles that have energy enough to pass through the $\rm \Delta E$ stage and Pulse Shape 
Discrimination for those stopping in the Si detectors. 
During the experiment identification thresholds of 0.5\,MeV for protons and 1\,MeV for 
$\rm \alpha$ particles were obtained.
The four PPAC modules were placed symmetrically with respect to the beam direction to 
measure evaporation residues. 
Each PPAC module consists of two coaxial PPACs (front and rear) operating in the same gas 
volume at the pressure of about 40\,Torr and mounted at the distance of 15\,cm from each other.
This gas pressure was sufficient to stop the ER between the two PPACs, and let the other
ions, like FF and elastic scattered beam particles, reach the rear PPAC. 
ERs are therefore selected by a vetoed Time of Flight technique: the time between the beam RF signal and 
the signal from the front PPAC is recorded only if the signal from the rear PPAC is missing.
This method allows to reject, with high efficiency, signals due to ions reaching the rear 
PPAC which are much faster and much lighter than evaporation residues. 
Each PPAC module was positioned at 4.5\,degrees with respect to the beam direction and subtended 
a solid angle of 0.8\,msr. 

Data were collected by requiring several triggering conditions, as described in 
\cite{Dini11}, to perform measurements of the single and coincidence yields in the same run. 
In this way an extended set of observables relative to the fusion-evaporation channel was 
obtained that consists of proton and $\rm \alpha$ particle multiplicities and energy spectra,
as well as angular correlations among the LCP and ER.
The angular distribution of evaporation residues was obtained by means of the 
electrostatic deflector PISOLO \cite{Begh85} in a complementary experiment as described in 
details in ref.\cite{Dini11}. 
The experimental data will be shown in section \ref{sec:Res}.

\section{Models}
\label{sec:Mod}
The extended set of observables, collected as described above, was compared with the predictions 
provided by the SM code PACE2$\rm \_$N11 and a dynamical model. 
The computer code PACE2$\rm \_$N11 is an extensively modified version of the code PACE2 
\cite{Gavr80} that simulates the multistep deexcitation of 
the compound nucleus both through light particle evaporation and fission. 
Light particle evaporation is implemented according to the Hauser-Feshbach formulation 
and the fission probability is calculated by using the transition-state model.
Fission barriers are computed with the Finite Range Liquid Drop Model (FRLDM) \cite{Sier86}. 
The competition between the different decay modes is treated with a Monte Carlo approach.
PACE2 was modified to implement options for leading parameters (transmission coefficients, 
level density parameter and yrast line) and to take into account the fission delay time 
$\rm \tau_{d}$ \cite{Dini11, Vard10, Moro12, Huyu16}. 
In particular, the fission decay width is given by the following formula:
\begin{equation}
	\Gamma_{f}=f(t)\Gamma_{BW}
\end{equation}
where $f(t)$ is a simple step function: $f(t)=0$  for $\rm t < \tau_{d}$ and $f(t)=1$ for $\rm t> \tau_{d}$. 
$\Gamma_{BW}$ is the Bohr-Wheeler width \cite{Bohr39}.

In the  dynamical model \cite{Nadt10, Vard15} the fission process is described with a 
stochastic approach code based on 3-D Langevin Equations combined with the computer code 
LILITA$\rm \_$N11 \cite{Gome81} to simulate the evaporation of LCP.
The nuclear shapes are described by (c, h, $\rm \alpha$) parametrization \cite{Brac72}, 
related to the collective variables which give the evolution of fissioning nuclei: (c) 
elongation, (h) neck size at given c and ($\rm \alpha$) mass asymmetry.
The use of LILITA$\rm \_$N11 allows to adopt the same options for leading parameters 
implemented in  PACE2$\rm \_$N11.

Both codes were modified in order to produce an event-by-event output of emitted ER and 
light particles that, filtered according to the response function of 8 $\rm \pi$LP, can 
be used for a direct comparison of predictions and experimental data.

\section{Results and Discussion}
\label{sec:Res}
The main goal of this analysis is to exploit the evaporation channel observables 
in order to put stringent constraints on the SM parameters included in the models used 
for fission dynamics studies.
These parameters are also used to describe the evaporation of light pre-scission particles 
emitted by the compound nucleus along the path from its formation up to the scission in 
two fragments.
This study will not solve the existing ambiguity concerning the ingredients describing the 
dynamical evolution of fissioning systems, as mass and friction \cite{Nadt16}, but at least 
can avoid that they will be affected by the incorrect definition of the SM parameters, as 
discussed in \cite{Vard15}.

In a previous work the $\rm ^{32}S + ^{100}Mo$ reaction at 200\,MeV extended data 
set was compared with SM calculations\cite{Dini11}.
In such work it was observed that if the analysis is limited to the pre-scission LCP 
multiplicities and FF cross section, as usually done in fission dynamics studies \cite{Paul94}, 
can be reasonably well reproduced without any delay.
From this result one could conclude that no dynamical effects take place in 
the decay of CN.
However different combination of the input parameters does not exclude the presence 
of a relatively small fission delay, as expected by the systematics \cite{Tho93}. 
On the other hand, a model, that strongly overestimates the ER particle multiplicities, 
cannot be considered reliable to estimate the fission time-scale through the pre-scission 
LCP multiplicities.
Therefore we used the ER angular distributions and LCP evaporative spectra to limit 
the SM parameters included in the dynamical model.
Calculations considering different dissipation mechanisms were compared with the 
full set of experimental data and most of the observables were well reproduced 
\cite{Nadt09,Vard15}. 
Therefore the ingredients describing the fission process were determined. 

In this section the data are presented in a way to progressively show the impact  
of SM paramters on different observables.
Finally the advantages coming by using a new detection system for ERs in operation
at LNL, will be illustrated in the light of model predictions.

The importance to explore very selective observables will be evidenced in the comparison 
with existing data and the importance of new observables will be illustrated by comparing
simulations filtered with the new detection system geometry.

\subsection{Evaporation residue angular distributions and evaporative LCP energy spectra}
\begin{figure}[h]
\resizebox{0.47\textwidth}{!}{
  \includegraphics{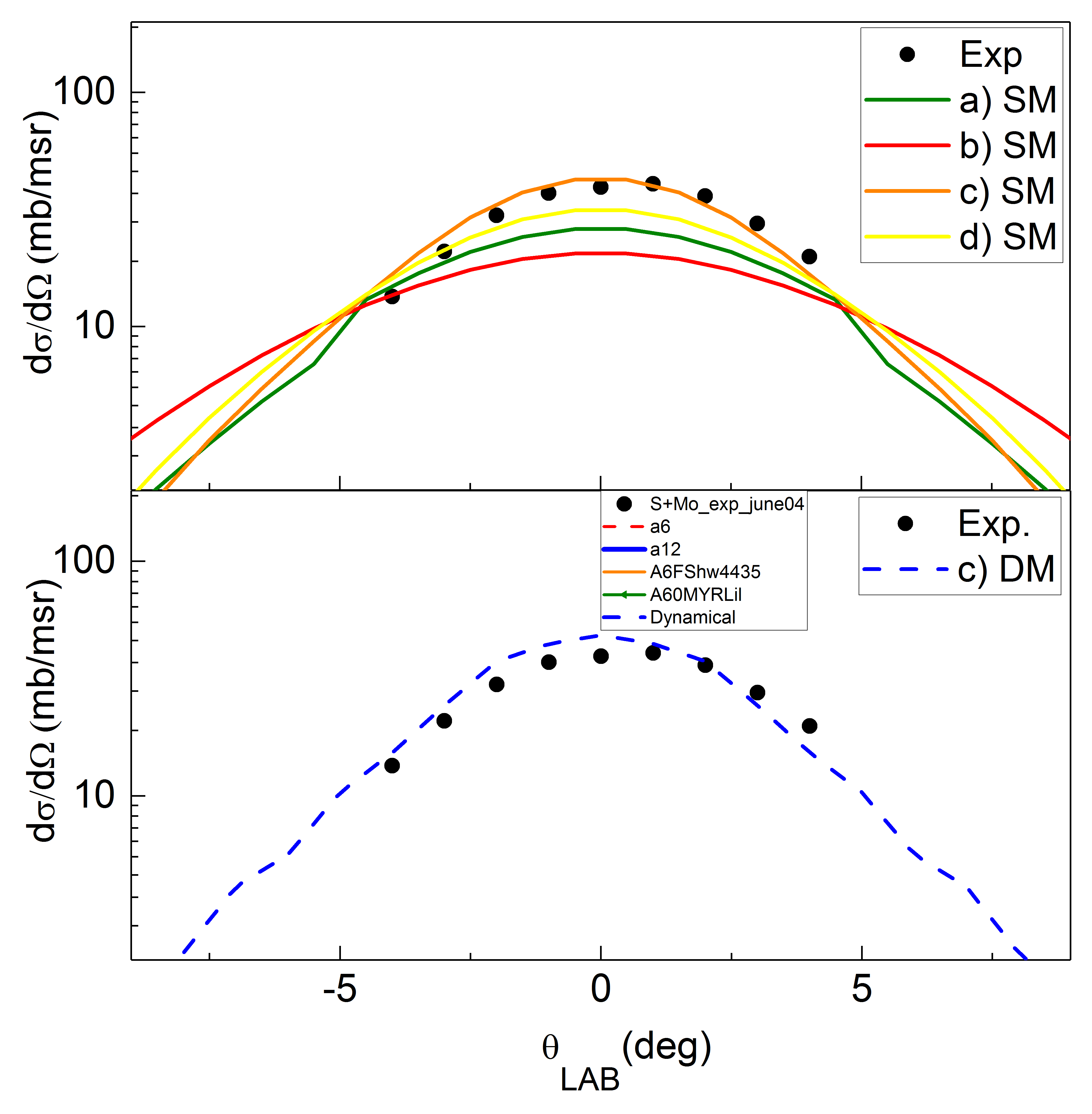}
}
\caption{(Color online) ER angular distribution of the reaction $\rm ^{32}S +^{100}Mo$. 
The experimental data (dots) are compared with the SM calculations (top) performed 
with the four prescription of Tab.~\ref{TabPRE} and with dynamical model 
calculation (bottom) performed by using only the prescription (c).
}
\label{ER_AD}
\end{figure}
In Fig.~\ref{ER_AD} we show the angular distribution of the evaporation residues 
compared with the results of the statistical model and the dynamical model calculations.
The experimental statistical errors are in the order of the point size.
By changing the leading parameters of the SM significant deviations are produced in the 
ER angular distributions as shown in Fig.~\ref{ER_AD}a.
The leading parameters of the simulations were the LCP transmission coefficients, 
the level density parameters $\rm a_{\nu}$ for particle evaporation, and the yrast line 
prescriptions. 
Transmission coefficients derived from the optical model (OM) \cite{Huiz61,Pere63, Wilm64} 
and from the fusion systematics (FS) \cite{Vaz84} in combination with a level density 
parameters between A/6 and A/12 were used.
The moment of inertia was calculated by adopting prescriptions for the yrast line with parameters 
from the Rotating Liquid Drop Model (RLDM) \cite{Sier86} or by assuming the compound 
nucleus as a the rigid sphere (RS) with $\rm r_{0}$= 1.2\,fm.

\begin{table}[h]
	\centering
		\begin{tabular}
			{|c|c|c|c|}
      \hline
      Prescriptions  & $a_{\nu}$      & Yrast Line  & Trans. Coef.  \\ 
      \hline
      a)       & A/6    & RLDM        & OM      \\      
      b)       & A/12   & RLDM        & OM      \\
      c)       & A/6    & RS          & FS      \\
			d)       & A/6    & RS          & OM      \\
      \hline
		\end{tabular}
	\caption{Prescriptions of SM parameter set adopted in the calculations 
	for 200\,MeV $\rm ^{32}S+^{100}Mo$ reaction.}
	\label{TabPRE}
\end{table}
The four different prescriptions used in the simulations shown in Fig.~\ref{ER_AD}a
are reported In Tab.~\ref{TabPRE}.
They were chosen among many combinations of the SM parameter values.
The aim was to explore the full range of variability of the observables under examination.
No fission delay was included in the calculations and the ratio $\rm a_{f} / a_{\nu}$ was 
kept equal to 1 due to relatively weak effects on the ER observables (see discussion in \cite{Dini11}).
The comparison of ER angular distributions in Fig.~\ref{ER_AD}a shows that the data are not 
well reproduced with a), b) and d) prescriptions, whereas a reasonable good agreement can 
be obtained by adopting the c) prescription.
The dynamical and the statistical models adopt different approaches to take into account 
fission-evaporation competition, however it was observed that, if the SM parameters of 
c) prescription are used, the data can be reproduced with comparable accuracy by both models,
as shown in Fig.~\ref{ER_AD}b for the DM.
Therefore the c) prescription was used in both models in the following comparisons with the
other experimental data.

\begin{figure}
\centering
\resizebox{0.3\textwidth}{!}{
\includegraphics{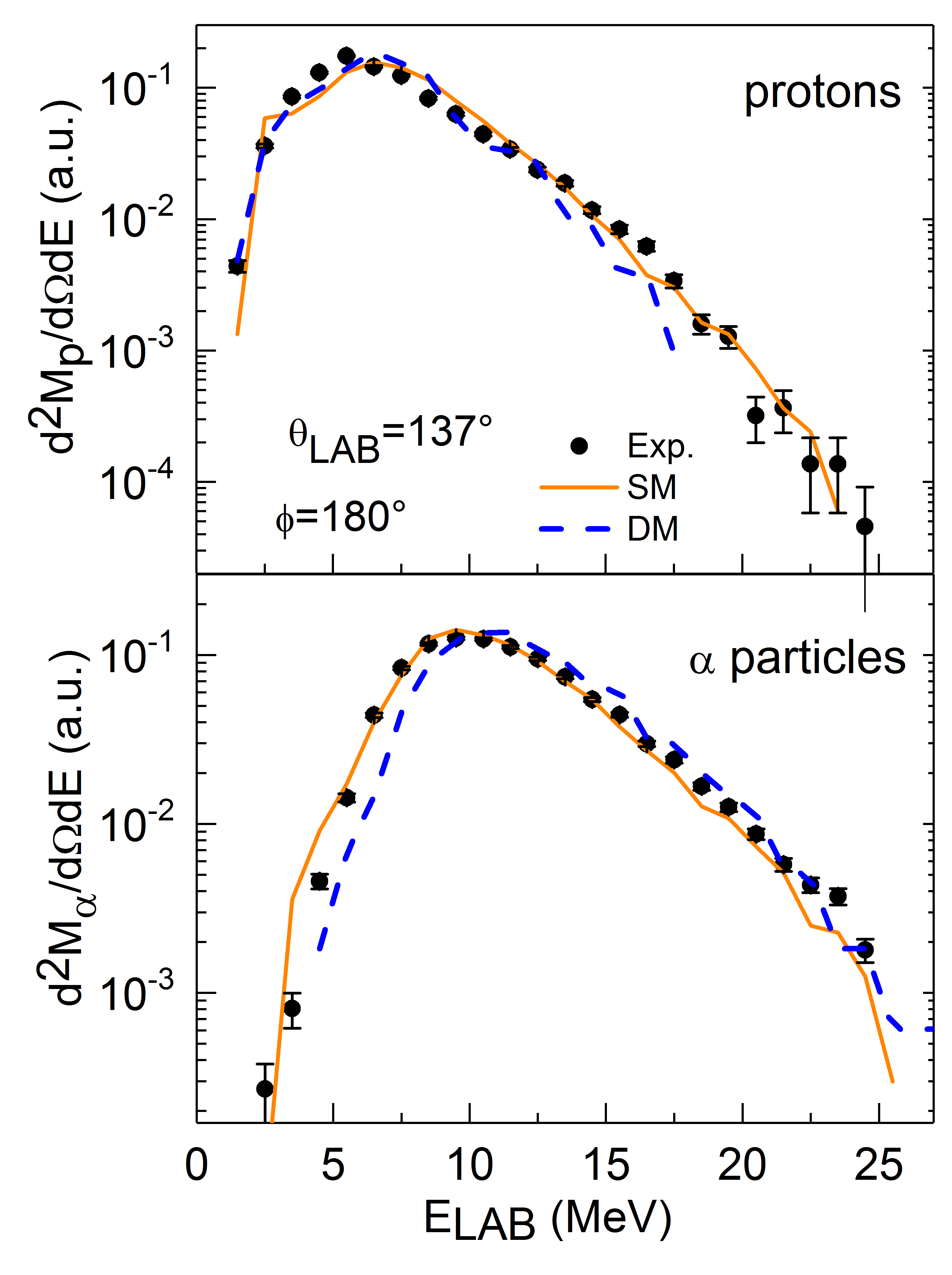}}
\caption{(Color online) Proton (top) and $\rm \alpha$ particle (bottom) energy spectra 
measured in coincidence with ERs. 
The experimental data (dots) are compared with the predictions of the statistical model 
(solid line) and the dynamical model (dashed line).
}
\label{SMspectra}
\end{figure}
In Fig.~\ref{SMspectra} the measured evaporative proton and $\rm \alpha$ 
particle energy spectra are compared with results of statistical and dynamical model simulations.
Both models very well reproduce the proton spectrum and the high energy side of 
$\rm \alpha$ particle spectrum, whereas the low energy side of this latter is better 
reproduced with the SM.
The same agreement holds also for the spectra measured at different angles.
Therefore it is reasonable to conclude that an overall agreement of energy spectra and 
angular distributions of evaporation residues can be obtained by adopting both the SM 
and the DM.
The use of the other prescriptions in Tab.~\ref{TabPRE} produce large deviations not only 
in the evaporation residues distribution, but also in the energy spectra (for details 
see \cite{Dini11}).
The sensitivity reachable in this comparison can provide indications for the most appropriate
SM parameters needed to describe the evaporation channel, but it is not sufficient to distinguish 
which model better reproduces the experimental data, being the observable so far considered 
only slightly influenced by the evaporation-fission competition.

\subsection{LCP multiplicities}
The LCP differential multiplicities were obtained by normalizing the 8$\rm \pi LP$ Ball-PPAC 
coincidence yields to the number of the ER events and then divided by the LCP detector solid 
angle.
\begin{figure} [h]
  \resizebox{0.35\textwidth}{!}{\includegraphics{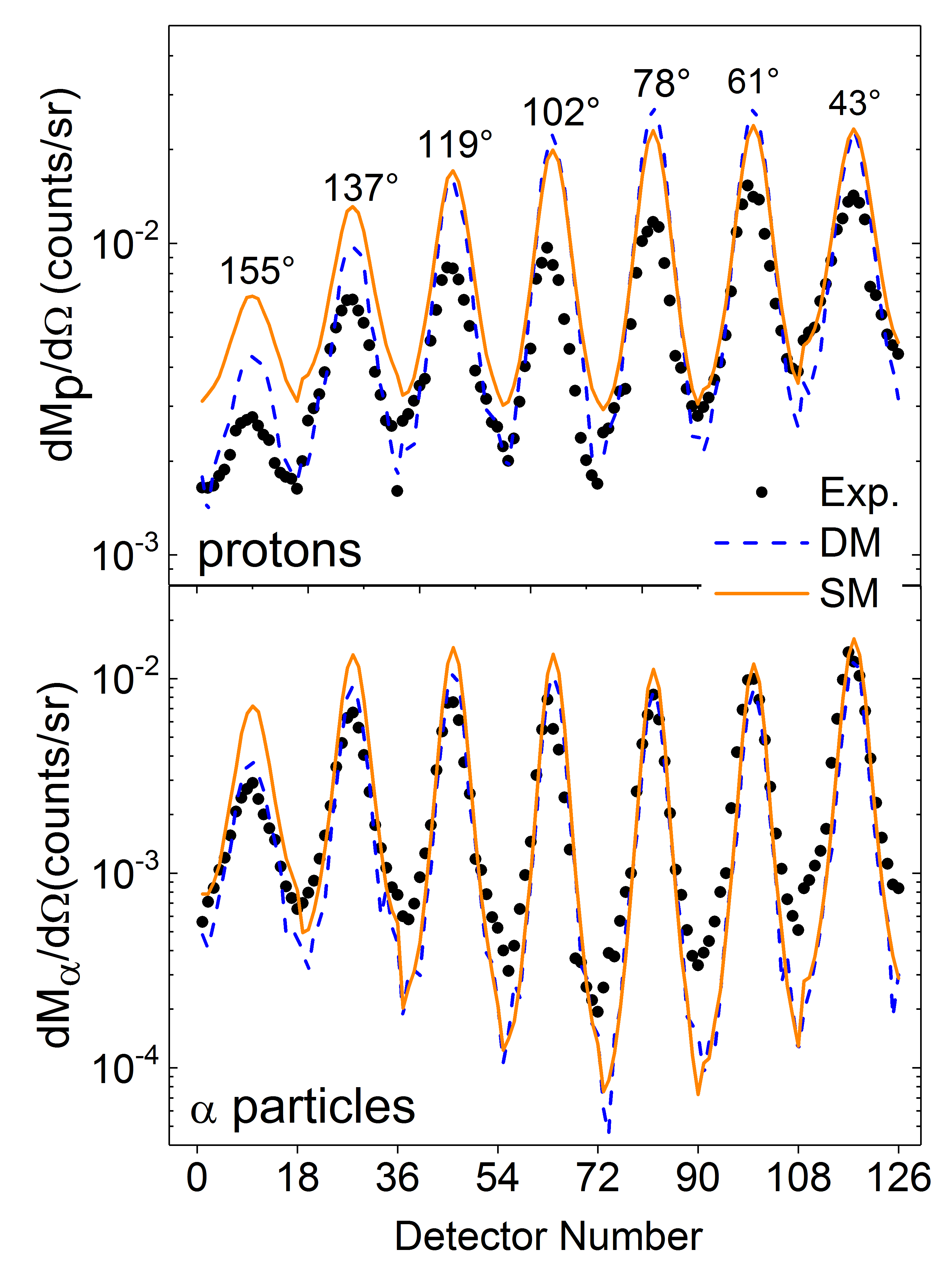}}
  \caption{(Color online) Experimental and calculated protons (top) and $\rm \alpha$ 
	particles (bottom)
	differential multiplicity distributions as function of  $\rm 8 \pi LP$ Ball detector 
	identification number. 	
  Black dots represent the experimental data, whereas the blue and orange lines are 
	the SM and DM simulations, respectively. 
	See the text for details.}
  \label{MA_cor}
\end{figure}
The resulting differential multiplicities of LCP are shown in Fig.~\ref{MA_cor} 
as a function of the Ball detector number, starting from the most backward angles. 
As mentioned before, for a fixed polar angle, the telescopes span the azimuthal angle 
from $\rm 0^{o}$ to $\rm 360^{o}$.
The oscillating behavior as function of the detector number is due to a combined effect of 
kinematics and of angular momentum of the composite system. 
In particular, the maxima correspond to events where ERs and LCPs are emitted in-plane 
and on opposite side with respect to the beam direction, $\rm \phi_{LCP} = 0^{o}$, whereas 
the minima occur when ERs and LCPs are emitted in-plane but on the same side with respect 
to the beam direction, i.e. $\rm \phi_{LCP} = 180^{o}$.
In the figure the mean polar angles corresponding to the telescope detecting the LCP,
$\rm \theta_{LCP}$, are indicated. 
By adopting different SM parameters the differential LCP multiplicities, and in particular those 
relative to the $\rm \alpha$ particle distributions, will be largely affected not only in terms 
of the bulk shapes ( see for instance the changing in the maxima to minima ratios in \cite{Dini11}), 
but also in terms of absolute values.
The experimental data are compared with the results of the SM and DM calculations 
assuming the c) prescription and without relative normalization.
The data seem to be better reproduced by the DM simulation being characterized by slightly 
smaller maxima.
Looking the data in Tab.~\ref{TabMULT} this is mainly true for protons.

\begin{table}
	\centering
		\begin{tabular}
			{|c|c|c|c|c|}
      \hline
             & $\rm M_{p}$          &$\rm M_{\alpha}$  & $\rm \sigma_{ER}$      & $\rm \sigma_{FF}$  \\ 
      \hline
      DM     & 1.20                 & 0.56                 & 793               & 143      \\
      SM     & 1.43                 & 0.72                 & 817               & 139              \\
      Exp.   & $\rm 0.90 \pm 0.14$  & $\rm 0.56 \pm 0.09$  & $\rm 828 \pm 50$  & $\rm 130 \pm 13$  \\
      \hline
		\end{tabular}
	\caption{The experimental and calculated mean particle multiplicities in the ER  
	channel together with the fission and evaporation cross sections.}
	\label{TabMULT}
\end{table}
By using the SM a consistent overestimation of both LCP multiplicities in the evaporation channel 
can be evidenced, while the evaporation and fission cross sections are well reproduced, see 
Tab.~\ref{TabMULT}.
The DM predicts not only evaporation and fission cross sections consistent with the experimental 
data, but also the evaporative $\rm \alpha$ particle multiplicity; only the proton multiplicity 
is slightly overestimate. 
Therefore by using the DM is possible to obtain a good overall agreement in the evaporation channel that
makes us confident on the use of such model for studying fission process \cite{Vard15}.

\subsection{New Observables}
High precision measurements of the evaporative LCP absolute multiplicities can be obtained 
only by measuring simultaneously ERs and LCPs emitted on the full solid angle.
However, the experimental apparatuses have limited angular coverages and only partial 
multiplicities are accessible.
For this reason the absolute multiplicities, as those reported in Tab.~\ref{TabMULT}, are not 
directly measured, but are extracted being driven by model predictions.
The use of 4 symmetric PPACs positioned at 4.5\,degrees around the beam direction is 
convenient because it allows to perform redundant measurements.
The redundancy is very useful because from one side assures the correct alignment of 
the beam impinging on the target, and from the other allows to apply the summing procedure 
to reduce statistical fluctuations.
By using our setup the partial multiplicities were obtained as the ratios among 
the yields of LCP in coincidence with ER's and the yields of all ER's detected by our PPAC's.
Afterward, the absolute multiplicities were estimated by considering the simulated 
ratios among the angle-integrated multiplicities and the quantity obtained by filtering 
the event by event output of the simulation code considering the 
geometry of the $\rm 8 \pi LP$ apparatus and the old PPAC system mounted at 4.5\,degrees.
The ratios obtained were $\rm  1.3 \pm 0.2$ for protons and $\rm 1.7 \pm 0.3$  for $\rm \alpha$ 
particles, therefore the reported errors take into account the wider variability produced 
by changing SM parameters in both PACE2$\_$N11 and LILITA$\_$N11 simulations.
Therefore to reduce the uncertainties and benchmark the fission models an experimental 
technique able to exclude the model dependence of observables is needed. 

The large angular coverage offered by the $\rm 8 \pi LP$ array can indeed favor the search 
for additional observables affected by the channel's competition in the CN decay process.
It was shown before that the reproduction of energy spectra and angular distributions of 
evaporation residues do not guarantee a good reproduction of the full de-excitation process. 
Because of this we extracted the differential multiplicities, which are simultaneously 
influenced by evaporation residues and LCP angular distributions and the competition with 
the fission channel over the extended $\rm 8 \pi LP$ angular range, and it was directly 
compare with the simulations in order to identify the most suitable SM parameters.
In order to further exploit this concept, we have built a new PPAC system to detect 
evaporation residues at the most significant forward angles for fusion reactions.

\begin{figure}
\resizebox{0.48\textwidth}{!}{%
  \includegraphics{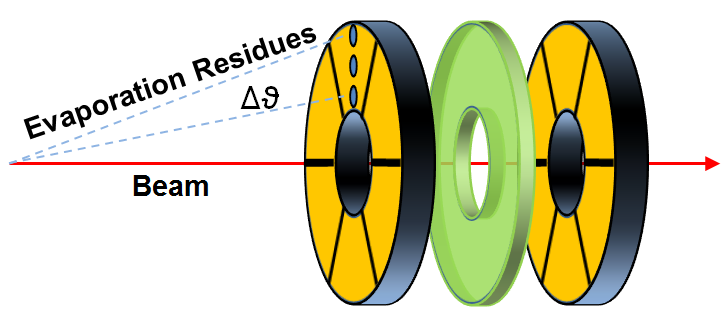}
}
\caption{Exploded view of the new PPAC system for evaporation residue detection.
This system can be combined with mask to measure ions emitted at around the forward 
direction at different polar angles, $\rm \theta_{ER}$, by spanning the 
$\rm \Delta \theta$ angular range.}
\label{PPAC}
\end{figure}
The new system for ER detection, whose schematic is shown in the Fig.~\ref{PPAC}, 
consists of two annular PPACs (front and rear) divided in 6 independent sectors 
with a wide area and an absorber foil mounted in between.
The absorber is adapted to the experimental conditions with thickness sufficient to stop 
only ERs and to let the other lighter ions, like LCPs and elastic scattered beam particles, 
passing through and reach the rear PPAC.
The ERs are therefore selected using as a veto the rear PPAC signal. 
In the previous experiment by using this system we were able to measure also the time of 
flight of ERs and improve the selectivity by excluding those produced by the beam interaction 
with target backing material.
To detect ER angular distribution between 3 and 8 degrees with respect to the beam a mask was 
mounted in front of the PPAC as indicated by the three blue spots in the Fig.~\ref{PPAC}.
Furthermore it should be underlined that thanks to the symmetric arrangement of the Ball 
telescopes and the PPAC sectors, also with this system it is possible to perform redundant 
measurements and get the consequent benefits.

\begin{figure}
  \centering
  \resizebox{0.43\textwidth}{!}{\includegraphics{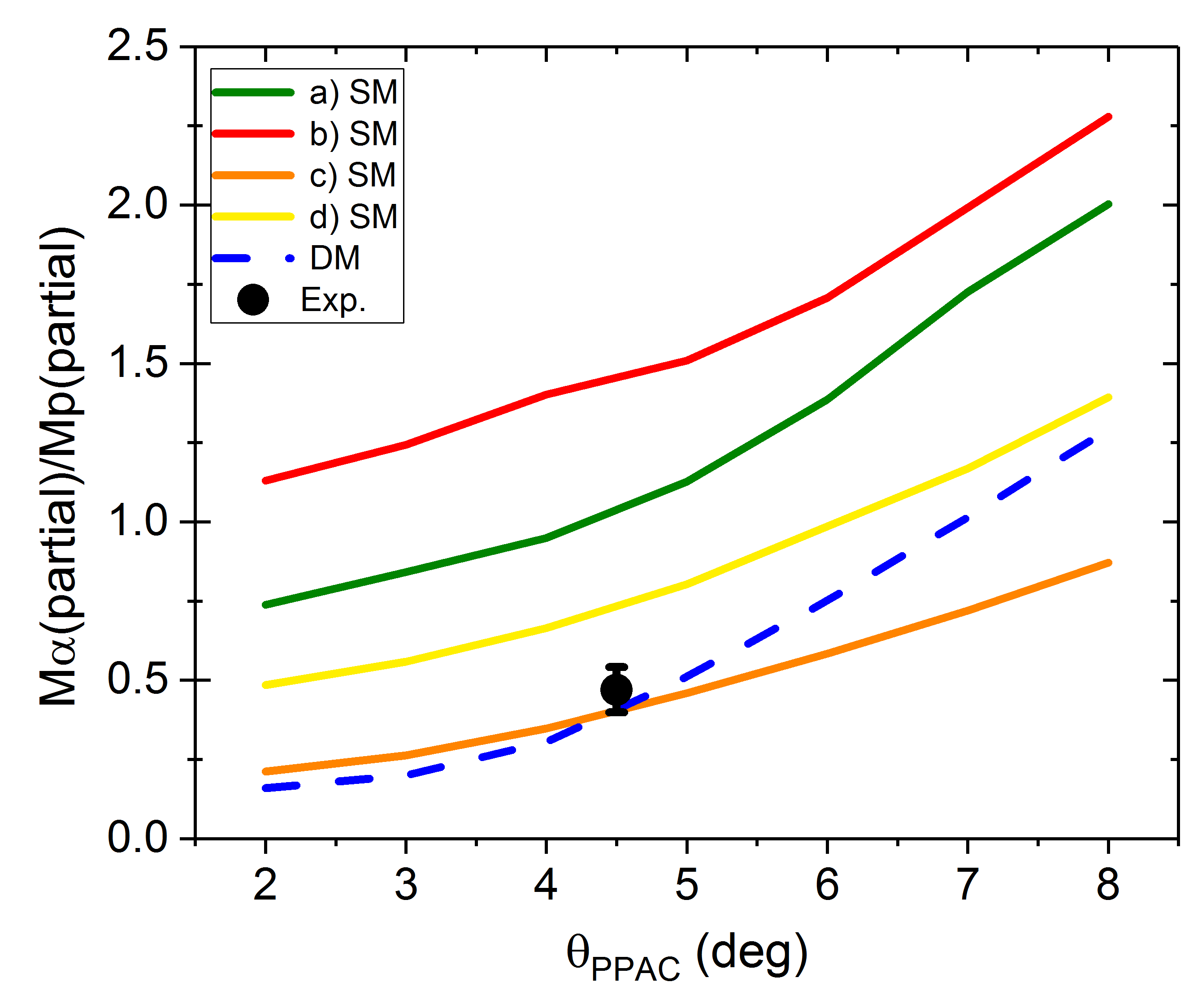}}
  \caption{(Color online) Ratios between $\rm \alpha$ particles and protons partial 
	multiplicities obtained with $\rm 8 \pi LP$ Ball filter as function of ER detection 
	angles. 
	The lines represent the SM calculations performed by using the four prescriptions of 
	Tab.~\ref{TabPRE} and the DM calculation, the dot is the experimental value measured
	with $\rm 8 \pi LP$ Ball and the PPAC at 4.5\,degrees.
	}
\label{Partial}
\end{figure}
The experimental point of the ratio between the $\rm \alpha$ particle and proton partial 
multiplicities measured with the old PPAC system mounted at 4.5\,degrees is shown in the 
Fig.~\ref{Partial}.
This quantity has the advantage to be at same time model independent and very sensitive 
not only to the decay kinematic, but also to the competition between evaporative LCP and 
fission channels.
Therefore we compared the experimental data point with simulations obtained by using the four 
prescriptions of Tab.~\ref{TabPRE} in PACE2$\_$N11 and the dynamical model.
In order to have a more general view of the behaviour, we simulates the trend of partial 
multiplicities by considering the different geometrical selections that can be applied 
by using the new PPAC, i.e. the measure of ER emitted at different mean angles ($\rm theta_{PPAC}$).
The ratios of partial multiplicities as function of $\rm theta_{PPAC}$ appear to be a 
good probe, as they are very sensitive to the SM parameters.
In general it is possible to observe a reduction of the proton partial multiplicities 
with the increase of $\rm \theta_{ER}$, on the contrary the $\rm \alpha$ particle 
multiplicities quickly increase.
This behaviour can be explained by considering that ERs at larger $\rm \theta_{ER}$ need 
more recoil that is preferably provided by $\rm \alpha$ particle evaporation.
Therefore the $\rm \alpha$ particle multiplicities increase.
However $\rm \alpha$ emission removes more excitation energy with respect to lighter 
particles, and the probability of a coincident proton emission will be reduced further 
increasing the ratio at high $\rm \theta_{ER}$.
More complicate is to explain the differences originated by each single SM parameters.
For a fixed yrast line, increasing $\rm a_{\nu}$ produces almost constant reduction 
in the ratios (see the prescriptions a) and b)).
Going from the yrast line calculated using the RLDM to the yrast line assuming the 
nucleus as a rigid sphere with $\rm r_{0}$= 1.2\,fm, that means a decrease of the 
moment of inertia, we observe a decrease in the ratios, that becomes larger with
$\rm theta_{PPAC}$ (see the prescriptions a) and d)).
While by changing transmission coefficients, i.e. passing from OM to FS ones, 
there is a decrease of about a factor two in the ratio that is almost 
independently from the ER angles considered, (see prescription c) and d)).
We observed that only combining all the effects producing a decrease in the 
$\rm \alpha$ particle to proton multiplicity ratios is possible to improve 
the reproduction of experimental data, and, in agreement with all the other 
observables, the prescription c) is the best set of parameters.
By comparing the statistical model calulcualtion with the dynamical one, only 
small differences are observed at $\rm theta_{PPAC}= 4.5\,degrees$ that cannot 
provide a conclusive indication about which of the two models better describe 
the CN decay.
However, the differences become larger and larger by increasing the 
$\rm \theta_{ER}$.
Therefore we planned to measure these observables as function of evaporation 
residues angles by coupling the $\rm 8 \pi LP$ detection apparatus and the new 
PPAC system.
Such very promising measurements highlight once more how the investigation of 
evaporation channel can play a relevant role in fission dynamic studies, in 
fact the reduction of the uncertainties on very exclusive observables can provide 
strong constraints not only to optimize the SM parameters, but also to validate 
the existing models.

\section{Summary and Conclusions}
\label{sec:Con}
The high charged particle multiplicities in pre-scission and evaporation channels existing 
in the intermediate fissility composite systems, makes them a good probe to get information 
on the fission process.
In this work we studied the evaporative light charged particles emitted by the $\rm ^{132}Ce$ 
nuclei at $\rm E_{x}$=122\,MeV by comparing the simulations with experimental data measured 
with the $\rm 8 \pi LP$ apparatus.
The simulations were performed using the SM code PACE2$\rm \_$N11 and a dynamical model based 
on the 3-D Langevin equations.

Evaporation residue angular distribution and LCP energy spectra were used to define the 
SM parameters included for SM and DM simulations.
By using the PACE2$\rm \_$N11 code, irrespective of the SM parameters adopted, the 
evaporative $\rm \alpha$ particle and proton multiplicities are overestimated.
Such a failure would affect also the emission probability in the pre-scission 
channel, making the extraction of the fission delay time unreliable.
Better results can be obtained by using the DM the evaporative LCP multiplicities 
which are reasonably well reproduced. 
This indicates that extended data set are essential to define simulations that can be  
considered reliable in order to address the open questions on fission dynamics.

Detailed studies evidenced the high sensitivity of LCP differential multiplicities 
not only to the SM parameters, but also to the nuclear shapes \cite{Dini16}.
The DM reproduces better also these observables, however not very large differences 
exist with respect to the SM predictions.
The main limitation to put stringent bounds on the SM parameters can be attributed 
to the uncertainties on experimental data, which are still slightly model dependent.
Therefore in order to increase the sensitivity we plan to measure in future experiments 
observables where particle and evaporation residue angular distributions are combined.
The possibility to use detection setup with a larger angular coverage clearly represent
a significant advance to benchmark the the fission models, and in conjunction with recent 
theoretical developments \cite{Nadt16} should act as a spur for future measurements in 
this field. 
By means of an annular PPAC the angular distribution of the ratios between the $\rm \alpha$ 
particle and proton partial multiplicities can be measured over an extended angular coverage.
This quantities have smaller uncertainties, being model independent, and according the 
predictions it is extremely sensitive not only to the main (leading) evaportive parameters, 
but also to the fission process description included in the statistical and the dynamical 
model adopted in this work.
In conclusion, the model strongly indicate that the exclusive observables of evaporation 
channel represent a powerful tool to define the fission dynamics that have to be considered 
in future studies.

\bibliography{ERtool}

\end{document}